\documentclass[reprint, amsmath, amssymb, aps, pra]{revtex4-2}

\usepackage{graphicx}
\graphicspath{{./Figures/}}
\usepackage{dcolumn}
\usepackage{bm}
\usepackage{mathrsfs}

\usepackage{float}

\usepackage{mathtools}
\DeclarePairedDelimiter\bra{\langle}{\rvert}
\DeclarePairedDelimiter\ket{\lvert}{\rangle}
\DeclarePairedDelimiterX\braket[2]{\langle}{\rangle}{#1\,\delimsize\vert\,\mathopen{}#2}

\begin{document}

\preprint{APS/123-QED}

\title{Quasi-analytical lineshape for the 1S-2S laser spectroscopy of antihydrogen and hydrogen}

\author{Levi O. A. Azevedo}
 \email{ldeazevedo@if.ufrj.br}
\author{Claudio Lenz Cesar}%
 \email{lenz@if.ufrj.br}
\affiliation{
 Instituto de F\'{i}sica, Universidade Federal do Rio de Janeiro, RJ, Brazil
}

\date{\today}

\begin{abstract}
  The accuracy of high precision and fundamental measurements of atomic transition frequencies via laser spectroscopy depends upon fitting the spectral data with a lineshape. With atomic hydrogen and antihydrogen 1S-2S two-photon spectroscopy, computer intensive Monte-Carlo simulations have been used to compute the optical Bloch equations in order to match and interpret the experimental spectra. For the highest resolutions, one tries to minimize saturation effects going to regimes of low excitation probability, where perturbation theory can provide reliable results. Here we describe an analytical approach to the lineshape based on perturbation theory accounting for the AC-Stark shift and ionization. The expressions can be used for beam experiments or integrated over the magnetic field profile for a trapped sample. Theses lineshapes, providing fast results, allow for studies of many systematic effects that influence the accuracy of the determination of the central frequency. This development has relevance to hydrogen beam experiments and to trapped hydrogen and antihydrogen, as developed by the ALPHA collaboration at CERN, for tests of the CPT-symmetry and the highest accuracy measurement on antimatter.

\begin{description}
    \item[Keywords]    
    Spectroscopy Lineshape, Hydrogen Spectroscopy, Antihydrogen, AC-Stark Effect, Magnetic Trap
\end{description}   
 
\end{abstract}

\maketitle

\section{\label{sec:level1}Introduction}

The 1S-2S transition frequency in hydrogen (H) is known with an accuracy of 4 parts in $10^{15}$~\cite{Hansch2011}. This transition in hydrogen, a simple and calculable atom, is the main determinant of the Rydberg constant~\cite{CODATA2018} and together with other constants and measurements set a stringent test of the theory of QED~\cite{KARSHENBOIM2019432}. Since the S-orbital penetrates the nucleus, this transition is also sensitive to the proton charge radius ($R_p$), object of intensive discussion~\cite{protonRadius} and best determined from muonic hydrogen spectroscopy~\cite{Pohl-ProtonRadius}. The proton charge radius causes a shift of 12~MHz in the 1S-2S frequency difference while the transition is presently determined to 10~Hz~\cite{Hansch2011}. Groups in Garching~\cite{Hansch2011}, Paris~\cite{LKB-Hspec-PhysRevLett.120.183001}, Colorado~\cite{Cooper_2023,Yost-PhysRevLett.130.203001}, and Zurich~\cite{Merkt-PhysRevLett.132.113001, Grasian} have been investigating hydrogen spectroscopy in a beam, while groups at Amsterdan~\cite{WalravenTrapPhysRevLett.70.2257}, MIT~\cite{MIT-PhysRevLett.77.255} and Turku~\cite{10.1063/5.0070037} have used ultracold trapped atoms which, in principle, allow for longer interaction times of the atoms with the laser beam. All experiments have not yet reached the 1S-2S natural linewidth limit, determined from the metastability of the 2S state with a decay time constant of ~121.5~ms~\cite{MIT-PhysRevLett.77.255,PhysRevA.69.052118}.

Changing to the antimatter world, antihydrogen ($\overline{\rm H}$) has been synthesized, trapped, and subjected to laser spectroscopy to parts in 10$^{12}$~\cite{1s2sAlpha} in the ALPHA experiment at CERN. The main motivation is to look for a violation of the Charge-Parity-Time (CPT) symmetry in search for an explanation of our matter-dominated Universe. While cold trapped atoms have possible advantages with respect to laser-atom interaction time, it suffers from complications due to the bias magnetic field, and the necessary field inhomogeneity of the trap. 

Either in traps or in beams, the required laser intensity -- given the two-photon nature of the transition with a large detunning from the one-photon virtual absorption -- induces a relevant AC Stark shift. Also, the laser photon, at 243~nm with 5.1~eV of energy, has sufficient energy to photoionize the 2S state. Thus, it is important to consider lineshape functions that account for these effects. In this manuscript we treat these effects and the inhomogeneous magnetic field in conditions typical of the ALPHA experiment~\cite{1s2sAlpha} with trapped laser cooled antihydrogen as obtained in 2023. 

This manuscript is organized as follows. A review of a simple lineshape from Ref.~\cite{Cesar_2016} and justification for using perturbation theory, in the conditions of the ALPHA experiment, is introduced in Sec.~2. In Sec.~3 we develop a treatment of the AC Stark and in Sec. 4 we include the 2S ionization. Both those treatments are compared to the optical Bloch equations (OBE) for different velocity classes in Sec. 5. In Sec. 6 we consider the spectra predicted with the influence of the magnetic field from the trap. In Sec. 7 we discuss deplenishing rates for the case of a finite trapped sample and differences in the excitation and detection processes. The manuscript closes with conclusions in Sec.~8.

\section{\label{sec:level2} Perturbation theory}

In this section, we recapitulate the expressions from second-order perturbation theory for the 1S-2S Doppler-free lineshape in Ref.~\cite{Cesar_2016}. We will disregard the natural decay linewidth as well as the momenta exchange between the laser beam intensity profile and the atom. A quantized motion formalism suitable for ultracold atoms -- not the case for this manuscript -- can be found in Refs.~\cite{PhysRevA.59.4564, PhysRevA.64.023418} where it is shown that this momenta transfer is the mechanism responsible for the "time-of-flight" spectrum, reconciling energy conservation.

\begin{figure}[b]
  \includegraphics[width=8.6cm]{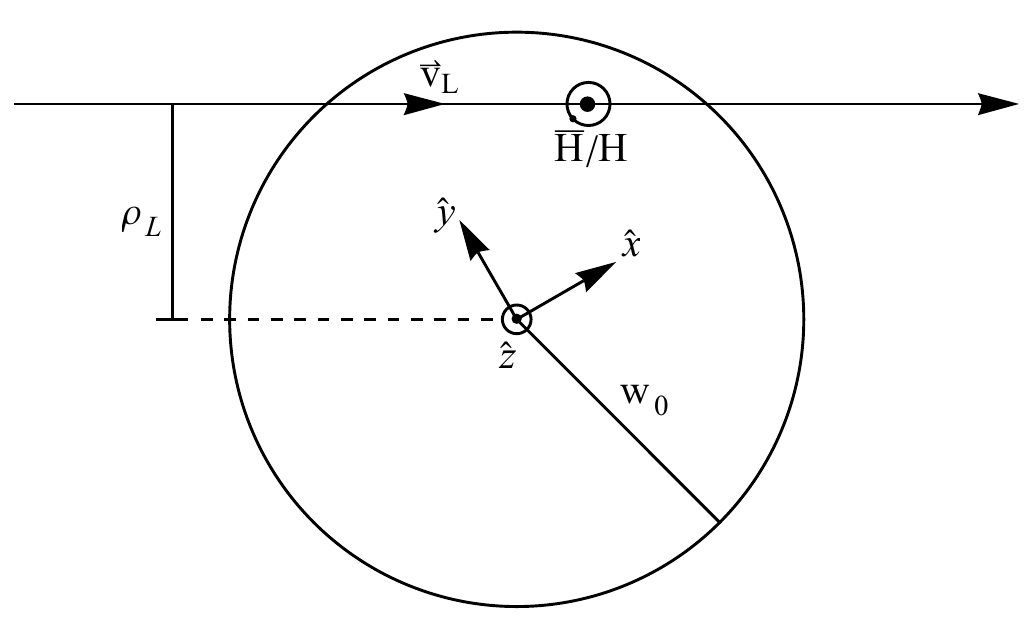}
  \caption{\label{fig:epsart} Illustration of a $\rm H$/$\overline{\rm H}$ transverse trajectory crossing the counterpropagating gaussian laser beams, along the z-axis,  at a beam waist w$_0$. The time of closest approach, at distance $\rho_{\rm L}$, is considered $t=0$ in our calculations. The atom's velocity is taken as constant along the crossing.}
\end{figure}

We consider two perfectly counterpropagating gaussian laser beams as in an optical cavity. The laser beams, propagating along the z-axis, have a local beam waist $\text{w}_0$, power $P$, and overlap perfectly. An $\rm H$/$\overline{\rm H}$ atom crosses the laser field with velocity $\vec{\text{v}}_L$, taken as constant during the crossing, and impact distance $\rho_{\rm L}$ at $t=0$ as shown in Fig.~\ref{fig:epsart}. After the atom, initially in the ground state, crosses the laser, the probability for it to get excited to the 2S state is given by (see Eq. A.7 of Ref.~\cite{Cesar_2016}):

\begin{equation} \label{eq:C2}
  |C(\delta\omega_{ca})|^2 = \gamma^2 \frac{P^2}{\text{v}_{\rm L}^2 \text{w}_0^2} e^{-\frac{4\rho{}_{L}^{2}}{\text{w}_0^2}} e^{-\frac{\text{w}_0^2\delta\omega_{ca}^2}{4 \text{v}_{\rm L}^2}} ,
\end{equation}
where $\delta\omega_{ca} = 2\omega_L-\omega_{ca}$ is the 2-photon detuning between the laser frequency, $\omega_L$, and the transition frequency, $\omega_{ca}$. The  factor $\gamma^2 = 1.362\times 10^{-7} \rm{m}^4 \, \rm{s}^{-2} \,\rm{W}^{-2}$ comes from the set of physical constants and summation of dipole moments that appear in the calculation as shown in Sec. 4. This value agrees with those of Refs.~\cite{HassJents, Gustafson_2021, PhysRevLett.39.1070}.

To obtain the excitation rate lineshape of the transition, one should integrate (\ref{eq:C2}) over the flux of atoms crossing the laser beam at a given position $z$ and all possible impact distances, i.e.,  $\delta\mathcal{L}(\delta\omega_{ca}) = (n_{\rm at}(z)\,\delta z) \int_{-\infty}^{\infty} \,d\rho_{\rm L}\int_{0}^{\infty} \,d\text{v}_{\rm L} \text{v}_{\rm L} f(\text{v}_{\rm L})\lvert C(\delta\omega_{ca})\rvert^2$. For a thermal sample with velocity distribution $f(\text{v}_{\rm L})\,d\text{v}_{\rm L}=2\frac{\text{v}_{\rm L}}{u^2} e^{-\frac{\text{v}_{\rm L}^2}{u^2}}\,d\text{v}_{\rm L}$ and $u=\sqrt{2\,k_B\,T/m}$ being the typical thermal speed, where $m$ is the atom's mass, $T$ is the sample temperature, and $n_{\rm at}(z)$ is the local atomic density, it results in

\begin{equation} \label{eq:LineshapePerturb}
 \delta \mathcal{L}(\delta\omega_{ca}) = 2.139\times10^{-7}\,  \frac{ P^2}{u\,\text{w}_0}e^{-\text{w}_0\lvert\delta\omega_{ca}\rvert/u}\, n_{\rm at}(z)\delta z \, .
\end{equation}

While $\text{w}_0$ and $\omega_{ca}$ are written as constants, they will be functions of the position along the laser beam taken as the z-axis. The gaussian laser beam divergence will vary the local beam waist, while for magnetically trapped samples the trap magnetic field will affect $n_{\rm at}(z)$ and the transition frequency $\omega_{ca}$.

The lineshape in Eq.~(\ref{eq:LineshapePerturb}) is the same as Biraben, {\it et al.}~\cite{birabenLineshape} in the limit that the natural linewidth is much smaller than the transit time, or time-of-flight, broadening. These expressions are not exact. The slowest crossing atoms will have a higher probability of excitation and ionization and may not be properly described by perturbation theory. These slow atoms influence mainly around the peak of the lineshape, as they have a longer transit time in the laser. These effects would smooth the peak of this cusped exponential lineshape while having little impact on its wings. 

\begin{figure}[H]
  \includegraphics[width=8.6cm]{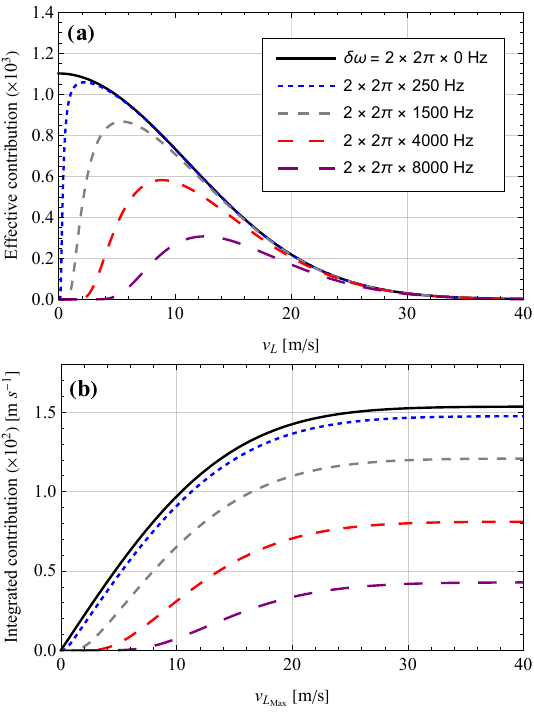}
  \caption{\label{fig:vLContribution} Effective (a) and integrated (b) contributions to the  spectrum at various detunings for different velocity classes. (a) is obtained from Eq.~(\ref{eq:C2}) multiplied by $2 (\text{v}_{\rm L}/u)^2 \exp[-(\text{v}_{\rm L}/u)^2]$ with $\rho_{\rm L}=0$, for a thermal sample at T=15~mK (u$\approx$15.7~m\,s$^{-1}$). (b) shows the integrated curve in (a) from $\text{v}_{\rm L}=0$ to $\text{v}_{\rm L}=\text{v}_{\rm L_{\rm MAX}}$. Notice that at 250~Hz detuning (blue dashed curve) at 243~nm, the integrated contribution to the spectrum by atoms with speeds below 1~m\,s$^{-1}$ is quite small, on the order of 4\% of its maximum value (when integrated from 0 to $\text{v}_{\rm L} \gg u$). The legends in (a) also apply in (b).}
\end{figure}

We show in Fig.~\ref{fig:vLContribution} that, at $ T \approx 15$~mK, the fraction of these slow atoms has a small contribution to the spectrum away from the peak frequency. Using Eq. (\ref{eq:C2}), for an atom crossing the laser beam with $P=200~$mW at a waist of 200~$\mu$m, impact distance $\rho_{\rm L}=0$ and speed of $1$~m\,s$^{-1}$ ($\approx u/15$), the excitation probability is 14\% at resonance. The number of atoms with speeds up to this value ($\text{v}_{\rm L} < u/15$) is $0.4\%$ of a sample at $15$~mK. For a sample of 2000 atoms, this would represent only 8 atoms at these low speeds and the excitation probability falls quadratically with the speed. This justifies why a perturbation theory approach should be able to produce accurate prediction for the excitation spectral lineshape under these conditions. In Fig.~\ref{fig:vLContribution}, the effective contribution of different velocity classes to absorption at different detunings is presented for a 15~mK sample. At each detuning, the overall absorption would be given by the integral over all the velocity classes. The contribution of very slow atoms ($\text{v}_{\rm L} < 1$~m\,s$^{-1}$) where Perturbation Theory fails, is small for detunings of 250~Hz (at 243~nm) or more. 

The optical Bloch equations for the system can be easily computed for specific atomic velocity classes, and thus, we will be able to compare the shapes and accuracy of our predictions. On the other hand, computing the excitation spectra for all the velocity classes, all impact parameters, and all positions along the laser beam is a task that requires a computer-intensive calculation and a Monte Carlo simulation if tracking the atoms through a trap. The present analytical perturbation theory method can be quickly integrated over the laser position and allows for invaluable studies of many systematic effects and parameters' scaling dependence.

\section{\label{sec:level3}AC-Stark effect}
The AC-Stark effect during the atom's crossing of the gaussian laser beam is a dynamical process, equivalent to a time varying phase shift.  Depending on the impact parameter, the shift will be different, being maximum for $\rho_{\rm L} = 0$. Here we consider an equivalent net effect for the crossing.

In the case of the 1S-2S transition $\rm H$/$\overline{\rm H}$, the AC-Stark effect on the atom crossing two counter-propagating gaussian laser beams with same geometric parameters and near the two-photons resonance ($\lambda \approx 243.133$ nm) is given by~\cite{HassJents}:

\begin{multline} \label{eq:ACShiftGaussian}
  \Delta\Omega_{\rm AC}(t) = 2\pi\times2\beta_{\rm AC}\,I(t) \\
  = \frac{8P\,\beta_{\rm AC}}{\text{w}_0^2}\exp\left[-\frac{2\rho_{\rm L}^2 + 2(\text{v}_{\rm L}t)^2}{\text{w}_0^2}\right],
\end{multline}
with~\cite{HassJents} $\beta_{\rm AC}=1.6771\times10^{-4} $~Hz (W/m$^2)^{-1}$ and $I(t)=\frac{2P}{\pi\,\text{w}_0^2}\exp\left[-\frac{2\rho_{\rm L}^2 + 2(\text{v}_{\rm L}t)^2}{\text{w}_0^2}\right]$ is the intensity profile~\cite{yariv} from a single laser beam seem by the atom along its trajectory.

\subsection{AC-Stark effect in dressed-picture}

We calculate the effect of this dynamic process on each atom using a "dressed-picture" approach, returning to the pertubation theory equations. Consider the states basis \{$\ket{a},\ket{b},\ket{c}$\}, where $\ket{a}\equiv\ket{1S}$ is the ground state, $\ket{c}\equiv\ket{2S}$ is the excited state, and $\ket{b}\equiv\ket{nP}$ represent all the intermediate states (including continuum) that are connected to $\ket{a}$ and $\ket{c}$ via electric dipole interaction. The arbitrary ket-state  $\ket{\Psi(t)}$, the $0^{\rm th}$-order Hamiltonian $H_0$ and the interaction Hamiltonian $V(t)$ can be written as:

\begin{subequations}
 \begin{equation} \label{eq:DressState}
    \ket{\Psi(t)} = 
    \begin{pmatrix}
      A(t)\\
      B(t)\\
      C(t)
    \end{pmatrix},
  \end{equation}
  \begin{equation} \label{eq:DressHamiltonian}
    H_0 = \hbar
    \begin{pmatrix}
      0 & 0 & 0\\
      0 & \omega_{ba} & 0\\
      0 & 0 & \omega_{ca}
    \end{pmatrix},
  \end{equation}
  \begin{equation} \label{eq:DressInteracHamiltonian}
    V(t) = 
    \begin{pmatrix}
      0 & -\hat{\vec{\mu}}_{op}\cdot\vec{E}(t) & 0\\
      -\hat{\vec{\mu}}_{op}\cdot\vec{E}(t) & 0 & -\hat{\vec{\mu}}_{op}\cdot\vec{E}(t)\\
      0 & -\hat{\vec{\mu}}_{op}\cdot\vec{E}(t) & 0
    \end{pmatrix}.
  \end{equation}
 
\end{subequations}
Here, $\hat{\vec{\mu}}_{op}$ is the electric dipole operator and $\vec{E}(t)$ is the electric field taken as:

 \begin{multline}
\label{eq:ELaserField}
  \vec{E}(t) = \frac{E_0\,\hat{\epsilon}}{2}\exp\left[-\frac{\rho_{\rm L}^2 + (\text{v}_{\rm L}t)^2}{\text{w}_0^2}\right]\times \\ 
  \left\{ \exp\left[i(k_1z-\omega_Lt)\right] +  \exp\left[i(-k_2z -\omega_Lt)\right]\right\} .
\end{multline}

In the above equation for the field, we have considered only the terms relevant for absorption (rotating wave approximation) and disregarded the other gaussian beam phases as we are only interested in two-counterpropagating photons absorption. $E_0$ represents the amplitude of the electric field, and the wavevectors $\vec{k_1}=-\vec{k_2}$ represent the two counterpropagating beams. We identify $\vec{E}(k_1,t)$ and $\vec{E}(k_2,t)$ with the expressions containing $k_1$ and $k_2$, respectively.

The AC-Stark shift affects both ground and excited states shifting their energy difference. We "dress" the ground state, 1S, with the laser field and insert on its energy the AC-Stark shift energy, writing a hamiltonian as:
\begin{equation} \label{eq:ACbraket}
  H_{AC}=-\hbar\:\Delta\Omega_{\rm AC}(t) \ket{a}\bra{a} ,
\end{equation}
where $\Omega_{\rm AC}(t)$ is given in (\ref{eq:ACShiftGaussian}). The negative sign appears since the AC-Stark effect increases the energy difference between the states.

Moving to the Interaction Picture on $H_0$ -- where $A(t)\rightarrow a(t)$ , $B(t)\rightarrow b(t)\:e^{-i\:\omega_{ba}t}$ and $C(t)\rightarrow c(t)\:e^{-i\:\omega_{ca}t}$ -- and applying  Schr\"{o}dinger's equation, one can obtain the $0^{th}$-order amplitude $a^{(0)}(t)$ as

\begin{equation} \label{eq:a0}
  a^{(0)}(t) = e^{i\frac{\sqrt{8\pi}P\:\beta_{\rm AC}}{\text{v}_{\rm L}\:\text{w}_0}\exp\left[-\frac{2\rho_{\rm L}^2}{\text{w}_0^2}\right] \left(1+{\rm Erf}(\frac{\sqrt{2} \text{v}_{\rm L}}{\text{w}_0} t) \right)},
\end{equation}
while $b^{(0)}(t) = 0$ and $c^{(0)}(t) = 0$. One can then proceed to higher orders in the E-field calculating $b^{(1)}(t)$ supposing absorption from beam 1 ($\vec{k}_1$) and $c^{(2)}(t)$ supposing absorption from beam 2 ($\vec{k}_2$) by solving the equations
\begin{equation} \label{eq:b1}
  \dot{b}^{(1)}(t) = \frac{\hat{\vec{\mu}}_{ba}\cdot\vec{E}(k_1,t)}{-i\:\hbar}e^{i\:\omega_{ba}t}a^{(0)}(t),
\end{equation}
and
\begin{equation} \label{eq:c2dot}
  \dot{c}^{(2)}(t) = \frac{\hat{\vec{\mu}}_{cb}\cdot\vec{E}(k_2,t)}{-i\:\hbar}e^{i\:(\omega_{ca}-\omega_{ba})t}b^{(1)}(t),
\end{equation}
which results in

\begin{widetext}
  \begin{multline} \label{eq:c2}
    c^{(2)}(t) = \left(\frac{E_0}{2\:i}\right)^2 \left(\frac{\hat{\vec{\mu}}_{cb}\cdot\vec{\epsilon}}{\hbar}\right)\left(\frac{\hat{\vec{\mu}}_{ba}\cdot\vec{\epsilon}}{\hbar}\right)e^{i(k_1-k_2)z}\int_{-\infty}^{t} \,dt_2\; e^{i(\omega_{ca}-\omega_L)t_2} e^{\frac{-\rho_{\rm L}^2}{\text{w}_0^2}}e^{-\frac{(\text{v}_{\rm L} t_2)^2}{\text{w}_0^2}}e^{-i\;\omega_{ba}t_2} \\ \int_{-\infty}^{t2} \,dt_1\;e^{i(\omega_{ba}-\omega_L)t_1} e^{\frac{-\rho_{\rm L}^2}{\text{w}_0^2}} e^{-\frac{(\text{v}_{\rm L} t_1)^2}{\text{w}_0^2}}e^{i\frac{\sqrt{8\pi}P\:\beta_{\rm AC}}{\text{v}_{\rm L}\:\text{w}_0}\exp\left[-\frac{2\rho_{\rm L}^2}{\text{w}_0^2}\right] \left(1+{\rm Erf}(\frac{\sqrt{2} \text{v}_{\rm L}}{\text{w}_0} t_1) \right)}.
 \end{multline}
\end{widetext}

The spatial phase in (\ref{eq:c2}) cancels, as $k_1-k_2=0$, resulting in zero momentum transfer or, alternatively, no Doppler shift. Given the enormous detuning from any one-photon transition with the intermediate state, the absorption of the two photons must occur essentially at the same time. This allows us to take the slow varying terms out of the integral in $dt_1$ by making $t_1\rightarrow t_2$ in those terms. The remaining integral in $dt_1$ with the necessary oscillating term for 2-photon absorption results in $e^{i(\omega_{ba}-\omega_L)t_2}/(i(\omega_{ba}-\omega_L))$. Using a linear polarization and calling $\hat{\vec{\mu}}_{cb}\cdot\vec{\epsilon} = \mu_{cb}$, and $\hat{\vec{\mu}}_{ba}\cdot\vec{\epsilon} = \mu_{ba}$, and substituting the electric field amplitude in terms of power ($P$) as $E_0^2 = 4 P/\pi\epsilon_0c\,\text{w}_0^2$ we obtain:

\begin{widetext}
  \begin{multline} \label{eq:c2b}
    c^{(2)}(t) = -\frac{P}{\pi\epsilon_0c\,\text{w}_0^2\,\hbar}  \frac{1}{i}\left(\frac{\mu_{cb}\mu_{ba}}{\hbar(\omega_{ba}-\omega_L)}\right)\;e^{-\frac{2\rho_{\rm L}^2}{\text{w}_0^2}}\;\int_{-\infty}^{t} \,dt_2\;e^{-i\left(\delta\omega_{ca}t_2-\frac{\sqrt{8\pi}P\:\beta_{\rm AC}}{\text{v}_{\rm L}\:\text{w}_0}\exp\left[-\frac{2\rho_{\rm L}^2}{\text{w}_0^2}\right] \left(1+{\rm Erf}(\frac{\sqrt{2} \text{v}_{\rm L}}{\text{w}_0} t_2) \right)  \right)}e^{-\frac{2(\text{v}_{\rm L}t_2)^2}{\text{w}_0^2}},
 \end{multline}
\end{widetext}
where, as above, $\delta\omega_{ca} = 2\omega_L-\omega_{ca}$ is the detuning from the two-photons transition. The expression for $c^{(2)}(t)$ should consider all the intermediate states $b$ and a factor of 2 from the exchange on the order of absorption of photons. Defining $\kappa_{cba}\equiv -\frac{1}{2h c \epsilon_0}\left(\frac{\mu_{cb}\mu_{ba}}{\hbar(\omega_{ba}-\omega_L)}\right)$, it reduces to:
\begin{widetext}
  \begin{equation} \label{eq:c2Final}
    c^{(2)}(t) = \frac{4\,P}{i\,\text{w}_0^2}\left(2\sum_{b}\kappa_{cba}\right)\;e^{-\frac{2\rho_{\rm L}^2}{\text{w}_0^2}}\;\int_{-\infty}^{t} \,dt_2\;e^{-i\left(\delta\omega_{ca}t_2-\frac{\sqrt{8\pi}P\:\beta_{\rm AC}}{\text{v}_{\rm L}\:\text{w}_0}\exp\left[-\frac{2\rho_{\rm L}^2}{\text{w}_0^2}\right] \left(1+{\rm Erf}(\frac{\sqrt{2} \text{v}_{\rm L}}{\text{w}_0} t_2) \right)  \right)}e^{-\frac{2(\text{v}_{\rm L}t_2)^2}{\text{w}_0^2}} .
  \end{equation}
\end{widetext}

For a detection process that measures the rate of excitation after the atoms finish crossing the laser, one needs to calculate $\lvert c^{(2)}(\infty)\rvert^2$, but the integral in (\ref{eq:c2Final}) has no analytical solution. A simple approximation of the time-dependent error function would yield ${\rm Erf}(\sqrt{2}\text{v}_{\rm L}t/\text{w}_0) \approx\sqrt{8/\pi} \,\text{v}_{\rm L}t/\text{w}_0$. Since it will be integrated in time as a phase, one can show that multiplying this term by $\sqrt{\pi/6}$ yields a better approximation as discussed in Appendix A. With this approximation it is possible to calculate $\lvert C_{\rm AC}\rvert^2 =\lvert c_{\rm approx}^{(2)}(\infty)\rvert^2$ analytically as

\begin{equation} \label{eq:C2AC}
  \lvert C_{\rm AC}(\delta\omega_{ca})\rvert^2=\gamma^2\frac{P^2}{\text{v}_{\rm L}^2 \text{w}_0^2}e^{-\frac{\text{w}_0^2}{4\text{v}_{\rm L}^2}\left(\delta\omega_{ca}-\Delta\omega_{\rm AC}(\rho_{\rm L})\right)^2}e^{-\frac{4\rho_{\rm L}^2}{\text{w}_0^2}} ,
\end{equation}
where $\gamma^2= \left(8\sqrt{\pi/2}\sum_{b}\kappa_{cba}  \right)^2$ with the value given above, after (\ref{eq:C2}). The summation term $\sum_{b}\kappa_{cba}$ can be identified as $\beta_{ge}=3.68111\times10^{-5}$~Hz (W/m$^2)^{-1}$ in Eq.~(E1) of Ref.~\cite{HassJents}. Thus, Eq.~\ref{eq:C2AC} is similar to Eq.~\ref{eq:C2} with the difference of an extra AC-Stark shift term in the detuning given by
\begin{equation} \label{eq:ACShiftBeforeRho}
  \Delta\omega_{\rm AC}(\rho_{\rm L}) = \sqrt{\frac{2\pi}{3}}\frac{4\,P\,\beta_{\rm AC}}{\text{w}_0^2}e^{-\frac{2\rho_{\rm L}^2}{\text{w}_0^2}}.
\end{equation}

As done in (\ref{eq:C2}) to obtain (\ref{eq:LineshapePerturb}), in order to calculate the excitation lineshape, for a slab of atoms at position $z$, one needs to integrate it over $\text{v}_{\rm L} f(\text{v}_{\rm L}) d\text{v}_{\rm L} \,d\rho{}_L$. Again, the integral of (\ref{eq:C2AC})  over $d\rho_{\rm L}$ has no analytical solution due to the term $\exp(-2\rho_{\rm L}^2/\text{w}_0^2)$ in $\Delta\omega_{\rm AC}(\rho_{\rm L})$. 

The interval of relevant values of $\rho_{\rm L}$ is defined by the external gaussian term $\exp(-4\rho_{\rm L}^2/\text{w}_0^2)$ in $\lvert C_{\rm AC}(\delta\omega_{ca})\rvert^2$. This corresponds to the power squared profile that appears from the two-photons nature of the transition. We can average $\Omega_{\rm AC}(\rho_{\rm L})$ over this envelope, obtaining an average AC-Stark shift for the laser profile
\begin{equation}\label{eq:omegaACmean}
  \overline{\Delta\omega_{\rm AC}} = \frac{8\,\sqrt{\pi}P\,\beta_{\rm AC}}{3\,\text{w}_0^2}\,.
\end{equation}

The final absorption lineshape, after the above approximation and integrating over $(n_{\rm at}(z) dz)\text{v}_{\rm L}f(\text{v}_{\rm L})\,d\text{v}_{\rm L}\,d\rho_{\rm L}$, is given by:
\begin{equation} \label{eq:LineshapePerturbAC}
   \delta\mathcal{L}(\delta\omega_{ca}) =\frac{\pi}{2}\gamma^2 \frac{P^2}{u\,\text{w}_0}e^{-\text{w}_0\lvert\delta\omega_{ca}-\overline{\Delta\omega_{\rm AC}}\rvert/u} \, n_{\rm at}(z) \delta z \,.
\end{equation}

We defer the comparison of the expressions above to the optical Bloch equations to after the treatment of ionization below.

\section{\label{sec:level4} Including Ionization}

In this section, we include ionization by adding a term $-(\gamma_i(t)/2) c(t)$ to the right-hand-side of Eq.~\ref{eq:c2dot}. We start by just considering the ionization rate part of the differential equation
\begin{equation} \label{eq:ci}
  \dot{c_i}(t) = - (\gamma_i(t)/2) c_i(t) \, ,
\end{equation}

with the ionization rate, which is time and laser intensity dependent, given by

\begin{eqnarray} \label{eq:gammai}
      \gamma_i(t) = 2\pi \beta_i 2 \frac{2P}{\pi \text{w}_0^2}\exp\left[- \frac{2 \rho^2}{\text{w}_0^2}\right]\exp\left[ -\frac{2 (\text{v}_{\rm L} t)^2}{\text{w}_0^2}\right] \nonumber \\
 = \kappa_{i} \exp\left[ -\frac{2 (\text{v}_{\rm L} t)^2}{\text{w}_0^2}\right] ,
\end{eqnarray} 
and where $\kappa_{i} \equiv 2\pi\beta_i 2 (2P/(\pi{\rm w}_0^2)) \exp\left[-2\rho^2 / \text{w}_0^2\right]$ and~\cite{HassJents} $\beta_i=1.202\times10^{-4}$~Hz$\,$(W/m$^2)^{-1}$. The solution of (\ref{eq:ci}) contains the time dependency: $\exp[-\kappa_{i} \text{w}_0 \sqrt{\pi/2}{\rm Erf}(\sqrt{2} \text{v}_{\rm L} t/\text{w}_0 )/(4{\rm v}_{\rm L}) ]$.

We incorporate this time dependence on (\ref{eq:c2dot}) by changing variables as 

\begin{equation} \label{eq:c2i}
c(t) = c_2(t) e^{-\frac{\kappa_{i}\text{w}_0 \sqrt{\pi/2}}{4 \text{v}_{\rm L}}\left( 1+ {\rm Erf}\left(\frac{\sqrt{2} \text{v}_{\rm L} t}{\text{w}_0} \right)\right)}.
\end{equation}

Changing to variable $c_2(t)$ in (\ref{eq:c2dot}) and adding  (\ref{eq:ci}) results in a time dependent integral given by:

\begin{widetext}
  \begin{equation} \label{eq:c2it}
    c_2(t) \propto \int_{-\infty}^{t} \,dt_2\;e^{-i\left(\delta\omega_{ca}t_2-\frac{\sqrt{8\pi}P\:\beta_{\rm AC}}{\text{v}_{\rm L}\:\text{w}_0}\exp\left[-\frac{2\rho_{\rm L}^2}{\text{w}_0^2}\right] \left(1+\frac{2/\sqrt{3} vL}{w0}t_2  \right)  \right)}e^{-\frac{2(\text{v}_{\rm L}t_2)^2}{\text{w}_0^2}} e^{+\frac{\kappa_{i} \sqrt{\pi/2} \text{w}_0 \left[ 1+ {\rm Erf}\left(\sqrt{2} \text{v}_{\rm L} t_2/\text{w}_0 \right)\right]}{4 \text{v}_{\rm L}} }  \end{equation}
\end{widetext}

Analyzing the real exponents, i.e., excluding the phase time dependence (energy conservation), we can obtain an analytical solution with reasonable accuracy to this integral by approximating ${\rm Erf}\left(\sqrt{2} \text{v}_{\rm L} t/\text{w}_0 \right) \approx \alpha\left(\sqrt{8/\pi} \,\text{v}_{\rm L} t/\text{w}_0 \right)$ above, with $\alpha=\sqrt{\pi/6}$, as done for the integral in (\ref{eq:c2Final}). With the analytical result of the integral (\ref{eq:c2it}) above, we can change the variable back to $c(t)$ from Eq.~\ref{eq:c2i} and take the limit $t \rightarrow \infty$ (full crossing of the laser beam). Squaring the wavefunction amplitude of probability, we obtain

\begin{widetext}
  \begin{equation}\label{eq:C2ACion}
  \lvert C_{\rm ACion}(\delta\omega_{ca})\rvert^2=\frac{\gamma^2P^2}{\text{v}_{\rm L}^2 \text{w}_0^2} e^{-\frac{4\rho_{\rm L}^2}{\text{w}_0^2}} e^{-\frac{\text{w}_0^2}{4\text{v}_{\rm L}^2}\left(\delta\omega_{ca}-\Delta\omega_{\rm AC}(\rho_{\rm L})\right)^2}  e^{-2 \left(\frac{\sqrt{2 \pi} \beta_i P e^{-\frac{2\rho_{\rm L}^2}{\text{w}_0^2}}}{\text{v}_{\rm L} \text{w}_0}\right)}  e^{\frac{1}{3}\left(\frac{\sqrt{2 \pi} \beta_i P e^{-\frac{2\rho_{\rm L}^2}{\text{w}_0^2}}}{\text{v}_{\rm L} \text{w}_0}\right)^2}
\end{equation}
\end{widetext}

The last term in (\ref{eq:C2ACion}) diverges as $\text{v}_{\rm L}\rightarrow{}0$. It can be written as $e^{(v_\text{c}/v_\text{L})^2}$, where the characteristic velocity $v_\text{c}=\sqrt{2 \pi/3} \, \beta_i P e^{-\frac{2\rho_{\rm L}^2}{\text{w}_0^2}} /\text{w}_0$. For  $P\approx200~$mW and $\text{w}_0\approx200~\mu$m, our typical conditions,   $v_\text{c} (\rho_{\rm L}=0) \approx 0.2~$m\,s$^{-1}$. For a thermal distributions at 10~mK there is negligible population at velocity classes below this value. Therefore, we neglect this term here, and, under this approximation, (\ref{eq:C2ACion}) becomes

\begin{widetext}
  \begin{equation}\label{eq:C2ACionApprox1}
  \lvert C_{\rm ACion}(\delta\omega_{ca})\rvert^2=\frac{\gamma^2\,P^2}{\text{v}_{\rm L}^2 \text{w}_0^2} e^{-\frac{\text{w}_0^2}{4\text{v}_{\rm L}^2}\left(\delta\omega_{ca}-\Delta\omega_{\rm AC}(\rho_{\rm L})\right)^2} e^{-\frac{4\rho_{\rm L}^2}{\text{w}_0^2}} e^{-\frac{\sqrt{8 \pi} \beta_i P e^{-\frac{2\rho_{\rm L}^2}{\text{w}_0^2}}}{\text{v}_{\rm L} \text{w}_0}}.
\end{equation}
\end{widetext}

The last term $e^{-(\sqrt{8 \pi} \beta_i P e^{-\frac{2\rho_{\rm L}^2}{\text{w}_0^2}})/(\text{v}_{\rm L} \text{w}_0)}$ decreases the excited state population, after crossing the laser beam, due to ionization. Atoms with low velocities will be more affected by ionization. On the other hand, for higher velocities, $\text{v}_{\rm L}\gg v_\text{c}$ , this term approaches unity and the amplitude of excitation at resonance will be dominated by $\gamma^2\,P^2 e^{-4\rho_{\rm L}^2/\text{w}_0^2}/(\text{v}_{\rm L}^2 \text{w}_0^2)$, from regular perturbation theory, as in (\ref{eq:C2}).

As in the steps between (\ref{eq:ACShiftBeforeRho}) and (\ref{eq:LineshapePerturbAC}), we need to integrate (\ref{eq:C2ACionApprox1}) over $\text{v}_{\rm L} f(\text{v}_{\rm L})  \,d\text{v}_{\rm L}\,d\rho_{\rm L} \, n_{\rm at}(z) dz$ in order to obtain the lineshape, but again the integral over $\text{v}_{\rm L} f(\text{v}_{\rm L})\,d\text{v}_{\rm L}$ has no analytical solution. To overcome this problem, we expand the ionization term $e^{-\text{v}_{\rm ion}/\text{v}_{\rm L}}$, with $\text{v}_{\rm ion}=(\sqrt{8 \pi} \beta_i P/ \text{w}_0) e^{-2\,\rho_{\rm L}^2/\text{w}_0^2}$ , in power series of $\text{v}_{\rm ion}/\text{v}_{\rm L}$ to second order as $e^{-\text{v}_{\rm ion}/\text{v}_{\rm L}}\approx1-(\text{v}_{\rm ion}/\text{v}_{\rm L})+(\text{v}_{\rm ion}/\text{v}_{\rm L})^2/2$. In order to avoid the singularity at $\text{v}_{\rm L}\rightarrow{}0$, we multiply the expansion by $\left(1-e^{-(\text{v}_{\rm L}/\text{v}_{\rm ion})^2}\right)$.  Under these approximations, (\ref{eq:C2ACionApprox1}) becomes

\begin{widetext}
  \begin{equation}\label{eq:C2ACionApprox2}
      \lvert C_{\rm ACion_{approx}}(\delta\omega_{ca})\rvert^2=\frac{\gamma^2\,P^2}{\text{v}_{\rm L}^2 \text{w}_0^2} e^{-\frac{\text{w}_0^2}{4\text{v}_{\rm L}^2}\left(\delta\omega_{ca}-\Delta\omega_{\rm AC}\right)^2} \left(1-\frac{\text{v}_{\rm ion}}{\text{v}_{\rm L}}+\frac{1}{2}(\frac{\text{v}_{\rm ion}}{\text{v}_{\rm L}})^2\right)\left(1-e^{-\left(\frac{\text{v}_{\rm L}}{\text{v}_{\rm ion}}\right)^2}\right)e^{-\frac{4\,\rho_{\rm L}^2}{\text{w}_0^2}},
\end{equation}
\end{widetext}
where both $\Delta\omega_{\rm AC}$ and $\text{v}_{\rm ion}$ depend on $\rho_{\rm L}$.

For integrating over $d\rho_{\rm L}$ we average $\Delta\omega_{\rm AC}$ and $\text{v}_{\rm ion}$ weighted by the external gaussian envelope $e^{-4\,\rho_{\rm L}^2/\text{w}_0^2}$, as done in (\ref{eq:omegaACmean}), resulting in $\overline{\text{v}_{\rm ion}}=\frac{4\sqrt{\pi} \beta_i P}{\sqrt{3}\, \text{w}_0}$. Integrating the result of this approximation on (\ref{eq:C2ACionApprox2}) over $n_{\rm at}(z)\delta z\, \text{v}_{\rm L} f(\text{v}_{\rm L})  \,d\text{v}_{\rm L}\,d\rho_{\rm L}$ we obtain the final result:

\begin{widetext}
  \begin{multline} \label{eq:lineshapeIoni}
  \delta\mathcal{L}_{\rm ion}(\delta\omega_{ca}) = \frac{\pi\,\gamma^2P^2}{2\,u\,\text{w}_0} \Bigg(  e^{-\text{w}_0\lvert\delta\omega_{ca}-\overline{\Delta\omega_{\rm AC}}\rvert/u}  -  \frac{\overline{\text{v}_{\rm ion}}}{u\sqrt{1+(\overline{\text{v}_{\rm ion}}/u)^2}}e^{-\text{w}_0\sqrt{1+(\overline{\text{v}_{\rm ion}}/u)^2}\,\lvert\delta\omega_{ca}-\overline{\Delta\omega_{\rm AC}}\rvert/\overline{\text{v}_{\rm ion}}}  \\  +\frac{\overline{\text{v}_{\rm ion}}^2}{\text{w}_0\,u\,\lvert\delta\omega_{ca}-\overline{\Delta\omega_{\rm AC}}\rvert}\left(e^{-\text{w}_0\lvert\delta\omega_{ca}-\overline{\Delta\omega_{\rm AC}}\rvert/u} - e^{-\text{w}_0\sqrt{1+(\overline{\text{v}_{\rm ion}}/u)^2}\,\lvert\delta\omega_{ca}-\overline{\Delta\omega_{\rm AC}}\rvert/\overline{\text{v}_{\rm ion}}}\right)  \\  -\frac{2\overline{\text{v}_{\rm ion}}}{\sqrt{\pi}\,u}\left( \mathcal{K}_{(0)}(\text{w}_0\lvert\delta\omega_{ca}-\overline{\Delta\omega_{\rm AC}}\rvert/u) - \mathcal{K}_{(0)}(\text{w}_0\sqrt{1+(\overline{\text{v}_{\rm ion}}/u)^2}\,\lvert\delta\omega_{ca}-\overline{\Delta\omega_{\rm AC}}\rvert/\overline{\text{v}_{\rm ion}}) \right) \Bigg) n_{\rm at}(z)\delta z,
 \end{multline}
\end{widetext}
where $\mathcal{K}_{(n)}(x)$ is the modified Bessel function of second kind and order $n$ ~\cite{Arfken}. Eq.~\ref{eq:lineshapeIoni} agrees with (\ref{eq:LineshapePerturbAC}) in the limit where there is no ionization $(\text{v}_{\rm ion}\rightarrow{}0$ or $\beta_i=0)$.

\section{Comparison to Optical Bloch Equations}

The optical Bloch equations offer a  precise way to describe the evolution of the atom's state as it crosses the laser beam, even at low velocities and higher powers where perturbation theory cannot be applied. On the other hand, it often does not allow the possibility of an analytical expression, since the coherences and populations depend on each other in a complex manner.

The result of the excited population for a certain velocity class and impact parameter can be calculated using the OBE. 
Using the notation $\rho_{ee}, \rho_{ge}, \rho_{gg}$, and $ \rho_{i}$ to represent, respectively, the density matrix elements of the excited state (2S), the coherence between the ground (1S) and excited state, the ground state, and the ionized population, we can solve the OBE by the following set of differential equations~\cite{HassJents,Rasmussen_2018}
\begin{subequations}
  \begin{equation} \label{eq:OBErhoee}
    \dot{\rho}_{ee}(t) = \Omega_L(\text{v}_{\rm L},\rho_{\rm L},t) \, \rm{Im}(\rho_{ge}(t))-\dot{\rho}_{i}(t)
  \end{equation}
  \begin{multline} \label{eq:OBErhoge}
    \dot{\rho}_{ge}(t) = -\frac{\gamma_i(\text{v}_{\rm L},\rho_{\rm L},t)}{2}\rho_{ge}(t) - i(\delta\omega_{ca}-\Omega_{\rm AC}(\text{v}_{\rm L},\rho_{\rm L},t))\\+\frac{i}{2}\Omega_L(\text{v}_{\rm L},\rho_{\rm L},t)(\rho_{gg}(t)-\rho_{ee}(t))
  \end{multline}
  \begin{equation} \label{eq:OBErhogg}
    \dot{\rho}_{gg}(t) = -\Omega_L(\text{v}_{\rm L},\rho_{\rm L},t)\, \rm{Im}(\rho_{ge}(t))
  \end{equation}
  \begin{equation} \label{eq:OBErhoi}
    \dot{\rho}_{i}(t) = \gamma_i(\text{v}_{\rm L},\rho_{\rm L},t)\rho_{ee}(t)
  \end{equation}
\end{subequations}
where
\begin{subequations}
  \begin{equation} \label{eq:OBEIniContions}
    \rho_{ee}(-\infty) = \rho_{ge}(-\infty) = \rho_{ion}(-\infty) = 0
  \end{equation}
  \begin{equation} \label{eq:OBEIniContions2}
    \rho_{gg}(-\infty) = 1
  \end{equation}
  \begin{equation} \label{eq:OBEOmegaL}
    \Omega_L(\text{v}_{\rm L},\rho_{\rm L},t) = 2\times2\pi\,\beta_{ge}2\frac{2P}{\pi\,\text{w}_0^2}e^{-\frac{2(\text{v}_{\rm L}t)^2}{\text{w}_0^2}}e^{-\frac{2\rho_{\rm L}^2}{\text{w}_0^2}} ,
  \end{equation}
  \end{subequations}

and where $ \Omega_{\rm AC}(\text{v}_{\rm L},\rho_{\rm L},t)$ and $ \gamma_i(\text{v}_{\rm L},\rho_{\rm L},t)$ are the same as in Eqs.(\ref{eq:ACShiftGaussian}) and (\ref{eq:gammai}) above.

In the above notation, as well as in the previous sections, the atom crosses the laser beam reaching its minimum distance from the laser axis at $t=0$. The natural decay from the 2S state into the 1S state is not considered here also, since the lifetime of the 2S state of 121.5~ms is much larger than the typical times the atoms take to cross the laser beam. Even for an atom with $0.1$ m/s, corresponding to $\sim 6\:\mu $K of energy, after crossing the laser, it would travel for 12 cm (typical 1/e) before decaying back to the 1S state. For $\rm H$ atoms at 1 mK only $0.06$ \% of the atoms would have velocity below 0.1 m/s crossing the laser.

\begin{figure}[H]
  \includegraphics[width=8.6cm]{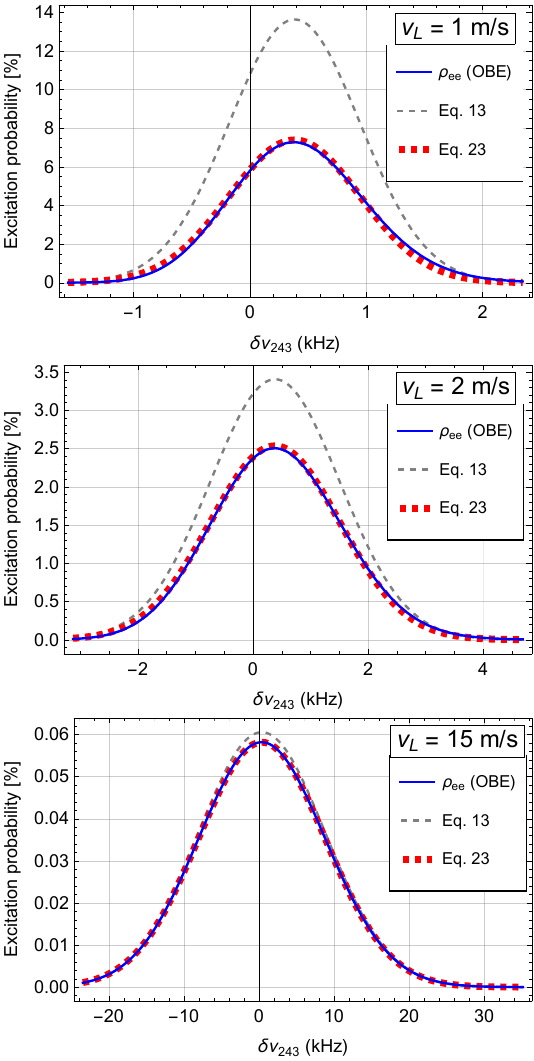}
  \caption{\label{fig:DirectComparisonOBE} Comparison between results from OBE (continuous blue line), $\lvert c^{(2)}(\infty)\rvert^2$ from Eq.~\ref{eq:C2AC} (without accounting for ionization, in dashed grey line), and Eq.~\ref{eq:C2ACionApprox2} (accounting for ionization, in dashed red line) as function of the laser detuning for different velocity classes. The laser parameters are $P=200$ mW and $\text{w}_0=200\:\mu$m. Notice the disagreement of neglecting ionization. With all the approximations, Eq.~(\ref{eq:C2ACionApprox2}) provides an excellent agreement to the OBE but there remains a small asymmetry (best noticed in the graph for $\text{v}_{\rm L}= 1$~m\,s$^{-1}$) which has a small effect on the accuracy of the central  frequency. See Appendix A and the text for discussion.}
\end{figure}

In order to test our Perturbation Theory expressions and approximations calculated above, we numerically solve the OBE for different velocity classes and compare to (\ref{eq:C2AC}), and (\ref{eq:C2ACionApprox2}). In Fig.~\ref{fig:DirectComparisonOBE} we see the comparison for three different speeds of 1, 2 and 15~m\,s$^{-1}$ with the atom crossing at $\rho_{\rm L}=0$ and laser parameters $P=200~$mW and $\text{w}_0=200~\mu$m. We see that even for low velocities, (\ref{eq:C2ACionApprox2}) agrees well with the OBE solution for $\rho_{ee}$ even at $\rho_{\rm L}=0$, the worst case. 
There remains a small asymmetry in the OBE result which is not captured by  (\ref{eq:C2ACionApprox2}), caused by the approximation discussed in Appendix A, which is relevant at low speeds and away from the resonance. This effect decreases at lower laser intensities, or higher $\rho_{\rm L}$, and higher $\text{v}_{\rm L}$.

\begin{figure}[b]
  \includegraphics[width=8.6cm]{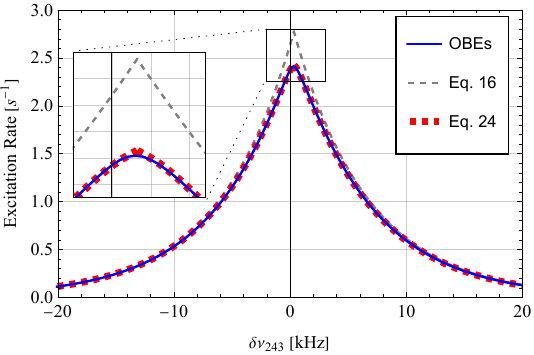}
  \caption{\label{fig:gasinabox} Excitation rate lineshape for a gas in a box. It shows a comparison of the results from the OBE, Eq.~\ref{eq:LineshapePerturbAC}, solution without considering ionization, and Eq.~\ref{eq:lineshapeIoni}, solution considering ionization. A zoom of the lineshape peak is shown in the inset. The parameters used were: $P = 200~$mW, $\text{w}_0 = 200~\mu$m, $T = 15~$mK and a total number of 10$^6$ atoms.}
\end{figure}

Considering typical conditions of a thermal sample -- in a thin box, non-diverging beam -- at 15~mK, $P$=200~mW and w$_0$=200~$\mu$m, we compared the OBE's results, by a summation over the velocity classes and different impact factors $\rho_{\rm L}$ with the proper weighting factors, with the analytical lineshapes coming from (\ref{eq:LineshapePerturbAC}) and (\ref{eq:lineshapeIoni}).  This comparison is shown in Fig.~\ref{fig:gasinabox}. By picking 11 points in the OBE lineshapes and fitting them with (\ref{eq:lineshapeIoni}) we estimate an error of less than 2 parts in 10$^{14}$  in the central frequency determination caused by this asymmetry. If a higher accuracy is needed, one direction is to use lower laser powers as discussed in the validity conditions in the last sections. Another alternative, is to perform the full numerical time integration of (\ref{eq:c2Final}) adding the ionization terms from (\ref{eq:c2i}) and (\ref{eq:c2it}) : 

\begin{widetext}
  \begin{multline} \label{eq:c2numerical}
    c^{(2)}(\infty) = e^{-\frac{1}{6}\left(\frac{\sqrt{2 \pi} \beta_i\,P\,e^{-\frac{2\rho_{\rm L}^2}{\text{w}_0^2}}}{\text{v}_{\rm L} \text{w}_0}\right)^2}  \frac{4\,P}{i\,\text{w}_0^2}\left(2\sum_{b}\kappa_{cba}\right)\;e^{-\frac{2\rho_{\rm L}^2}{\text{w}_0^2}}\; e^{-\frac{\kappa_{i}\text{w}_0 \sqrt{\pi/2}}{4 \text{v}_{\rm L}} 2}   \\ \times \, \int_{-\infty}^{\infty} \,dt_2\;e^{-i\left(\delta\omega_{ca}t_2-\frac{\sqrt{8\pi}P\:\beta_{\rm AC}}{\text{v}_{\rm L}\:\text{w}_0}\exp\left[-\frac{2\rho_{\rm L}^2}{\text{w}_0^2}\right] \left(1+{\rm Erf}(\frac{\sqrt{2} \text{v}_{\rm L}}{\text{w}_0} t_2) \right)  \right)}e^{-\frac{2(\text{v}_{\rm L}t_2)^2}{\text{w}_0^2}} e^{+\frac{\kappa_{i}\text{w}_0 \sqrt{\pi/2}}{4 \text{v}_{\rm L}}\left( 1+ {\rm Erf}\left(\frac{\sqrt{2} \text{v}_{\rm L} }{\text{w}_0} t_2 \right)\right)} ,
   \end{multline}
\end{widetext}
where the first (exponential) term was introduced to cancel the higher order ionization term that diverges as $\text{v}_\text{L} \rightarrow 0$ as discussed after Eq.~\ref{eq:C2ACion}. For numerical integration, in practice, one can take the time limit to $\pm 3 {\text{w}_0}/\text{v}_{\rm L} $ instead of $\pm\infty$. Once the time integral is performed, one takes the absolute squared and then proceed to numerical integrations over $n_{\rm at}(z)\delta z\, \text{v}_{\rm L} f(\text{v}_{\rm L})  \,d\text{v}_{\rm L}\,d\rho_{\rm L}$. 




\section{\label{sec:level5} Magnetic trap}

In this section we discuss spectra generated by the integration of (\ref{eq:lineshapeIoni}) along the laser beam, the $z$--axis, with a magnetic field varying in space. This integration results in an excitation rate, in s$^{-1}$, for each frequency detuning.  The gaussian laser beam divergence can be taken into account by substituting $\text{w}_0$ by $\text{w}(z)=\text{w}_0\sqrt{1+(z/z_R)^2}$, where $z_R=\pi\,\text{w}_0^2/\lambda$ is the Rayleigh length. While we are taking ionization into account, the results obtained are for atoms found in the excited state after crossing the laser beam, and not ionized atoms. 

In the case of samples trapped in a magnetic field one must include the local atomic density variation and the frequency shift caused by the Zeeman, $\delta \omega_{\rm Zee}(B) \approx 2\pi\times$186~kHz $B$/(1 T), and Diamagnetic, $\delta \omega_{\rm Dia}(B) \approx 2\pi\times$388~kHz $B^2$/(1 T)$^2$, effects~\cite{1s2sAlpha}.  For the analytical lineshape we consider the atomic transition frequency at the magnetic field where the atom crosses the laser beam at closest approach to the laser.  This is equivalent to taking an average of the transition frequency in the short trajectory, where these values change linearly to first order. This approximation also needs validation for the typical conditions considered, as discussed below.  In Appendix B, a solution of the equations above accounting for this linearized change in frequency, which can be integrated analytically in time ($t_2$) but would require numerical integration over the other variables ($ \text{v}_L, \rho_L$ and $z$), is presented. 


For trapped atoms, or by considering the laser divergence, the final integral over $n_{\rm at}(z) d z$ has to be solved numerically. For a thermal sample, the atomic distribution is $n_{\rm at}(r,z) \propto \exp(-\mathcal{U}(r,z)/k_B\,T)$, where $\mathcal{U}(r,z)\approx\mu_B\,\lvert B(r,z)\rvert$ is the potential energy of a H/$\overline{{\text H}}$ atom in the ``1S$_d$" hyperfine state at position $z$. Using a program like Mathematica~\cite{mathematica}, where our calculations were performed, this integral can be evaluated inside a nonlinear fitting routine to adjust parameters such as the central frequency, effective temperature, amplitude and possible offset (coming from "dark" counts) and produce results in seconds. 

\begin{figure}[h]
  \includegraphics[width=8.6cm]{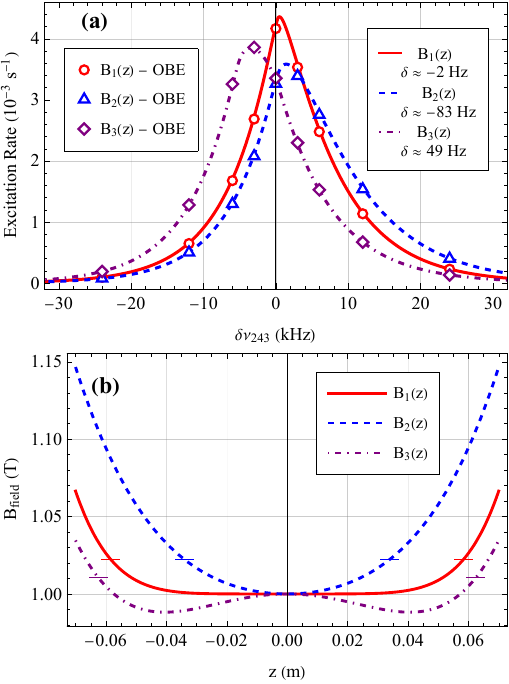}
  \caption{\label{fig:CompareLineshapes} Excitation rate lineshape fits in three different magnetic traps. The spectra (a) for different magnetic field configurations (b) along the laser beam are shown. The rates in (a) are normalized to a single atom sample with $T=15$~mK, $P=200$~mW and $w_0=200~\mu$m. The markers (segments) in (b) represent the magnetic field where the trapping energy equals the thermal energy. The lines asymmetries, in this case, are dominated by the magnetic field inhomogeneity locally shifting the 1S-2S transition frequency. The points in (a) are obtained by solving the OBE, considering the field seen by the atom in time, with a fine grid of $\text{v}_L, \text{v}_z, z_c$, and $\rho_L$. The lines are fits to the points using Eq.~\ref{eq:lineshapeIoni}, resulting in an error $\delta$ in the predicted center frequency shown in the figure. See text for further details.} 
\end{figure}

As examples of lineshapes and validation of the analytical lineshape, we use three idealized different trapping magnetic fields configurations as shown in Fig.~\ref{fig:CompareLineshapes}. The profiles of the magnetic fields along the $z$-axis are shown in Fig.~\ref{fig:CompareLineshapes}(b), and in Fig.~\ref{fig:CompareLineshapes}(a) the corresponding resulting spectra are presented. The radial confinement was taken as an ideal octupolar field, whose radial magnetic field is given by $B_r(r,z) = 1.1 (r/0.0225)^3~$T. The colored bars, or markers, in Fig.~\ref{fig:CompareLineshapes}(b) identify a magnetic depth corresponding to the thermal energy ($\mu B = k_B T$). The detuning $\delta\nu_{243}$ is relative to the frequency at $z=0$. The red curve represents a field of form $B_1(z)=((z/z_1)^6+1)\times 1~$T, the dashed blue curve $B_2(z)=((z/z_2)^2+(z/z_3)^4+1)\times 1~$T and the dot-dashed purple curve $B_3(z)=(-(z/z_4)^2+(z/z_5)^4+1)\times 1~$T, where $z_1=0.11~$m, $z_2=0.25~$m, $z_3=0.137~$m, $z_4=0.264~$m and $z_5=0.123~$m. The flatter field profile (continuous red line) exhibits a spectrum with less asymmetry and narrower linewidth than the other profiles. This occurs because most of the atoms interact with the laser while probing similar magnetic fields.

In Fig.~\ref{fig:CompareLineshapes}(a) the points are results of a lengthy computer calculation of the excited population by solving the OBE across the laser beam, in a time interval of $-2\text{w}_0/\text{v}_L$ to 2$\text{w}_0/\text{v}_L$ and considering  $\text{v}_L$ and $ \text{v}_z$ constant during the cross, and using a fine mesh in $ \text{v}_L \in (0.1,55)\,$m\,s$^{-1}$ with 183 values, $\rho_L  \in (-1.5,1.5)\, \text{w}_0$ with 120 values and $z  \in (-80,80)\,$mm with 80 values. The lineshapes are fits to these points from Eq.~\ref{eq:lineshapeIoni} integrated over $z$, with $\delta\omega_{ac}$ corrected for the local $B(z)$ field. Notice the errors in the fitted central frequency $\delta$, which, for the blue dashed line from $B_2(z)$, is $-2\times83$~Hz at the full 1S-2S frequency of $2.466\times10^{15}$~Hz, which is equivalent to $-7$ parts in 10$^{14}$, and only $-2$ parts in 10$^{15}$ for the flatter field $B_1(z)$. A negative(positive) error $\delta$ means that the analytical lineshape predicts a higher(lower) central frequency than the one used in the OBE.

The formalism can also be applied to non-thermal samples. A trapped thermalized and suddenly truncated sample (truncated Maxwell-Boltzmann) is easily implemented by integrating in $\text{v}_{\rm L}$ only up to a maximum $\text{v}_{\rm MAX}$. In the case of a thermalized truncated sample, one should use a different distribution such as Eq.~(20) in Ref.~\cite{Walraven}. A two-temperatures model, where the radial and axial temperatures are not the same during the spectroscopy exposure -- due to some anisotropic cooling faster than the ergodicity time -- can be trivially implemented. With a more general non-thermal distribution, such as an adiabatically cooled sample~\cite{Danielle2024}, one might need to perform an extra numerical integral in $\text{v}_{\rm L}$.

\section{\label{sec:level6} Discussion}

Our formalism predicts the 1S-2S excitation rates, or excitation spectrum, in the limit of a very large sample. Each experiment with hydrogen or antihydrogen uses different detection techniques which may or not correlate directly with the excitation rate. The detection process may add a complicated dynamical process that cannot be captured by our simple excitation rates and may only be computed via Monte-Carlo simulation. In some experiments, an applied electric field -- right after the excitation laser interacts for short times compared to the 2S decay time -- induces a quench, mixing the 2S with the 2P state that results in a quick Ly$_\alpha$ decay and detection. In this case the detected spectrum will faithfully reproduce the excitation spectrum.  In the case of the ALPHA experiment, the antihydrogen atoms had been usually detected by two processes: by ionization following excitation or by a spin-flip decay, both of which lead to detected annihilation of antiprotons on the walls~\cite{Rasmussen_2018}. The ionization process, for trapped atoms, could occur (i) in the same passage as the excitation of the atom through the 243~nm laser, or may occur in (ii) a subsequent passage of the excited through the laser beam. If process (ii) is dominant and the atoms happen to pass quickly (in times shorter than the 2S decay time of ~121.5~ms) again through the laser beam, then the detected spectrum will essentially mimic the excitation spectrum. In process (i), the detected spectrum will not be faithfully represented by the excitation spectrum and, in particular, it will feature a slightly wider lineshape. It is possible to obtain the ionized spectrum from the time dependence of the excitation probability by numerical integration. But, since the focus of this work is on analytical expressions, we do not present ionized spectra here. If the detection process involves a combination of these (i) and (ii) processes, and/or the spin-flip process, then one has to rely on a Monte-Carlo simulation to obtain the spectrum since the dynamics of the atom though the trap influences these processes. 

It is also relevant to mention that in traps, the atoms see an effective relativistic electric field by crossing the magnetic field and this adds a quenching field between the 2S and the 2P states. On the other hand, the relevant levels are much shifted away -- from the zero-field Lamb-shift -- in the fields typically used in the ALPHA experiment (see details in Ref.~\cite{Rasmussen_2018}). Thus, the decay time due to this process is also not relevant compared to the time the atoms take to cross the laser beam.

Another issue that affects the measured spectra is that of a finite sample, and its deplenishing. It is important to notice that the equations above, such as (\ref{eq:lineshapeIoni}), produce an excitation rate once integrated in $z$. In the case of the ALPHA experiment, the trapped sample is finite and the excitation followed by ionization or spin-flip to an untrapped state leads to sample loss, or deplenishing. One can normalize the excitation fraction, after a certain time integration, to the total number of atoms and predict the fraction of the total atoms that will be excited at each detuning. If all those excited atoms were lost for detection, one can directly compute the sample's deplenishing. If only a fraction of the excited atoms will be lost for detection while the other fraction will return to the ground state, then the actual deplenishing rate will be smaller than implied by the excitation rate.

Despite the complications discussed above, the  formalism presented here for the excitation spectrum should match well detected spectra in many different scenarios as long as the conditions for the approximations are satisfied. The validity of the approximations used can be summarized by relating speed, laser power and beam waist. For a  sample in which the number of atoms with speeds $\text{v}_{\rm L} < \overline{\text{v}_{\rm ion}} =4\sqrt{\pi/3} \, \beta_i P/ \text{w}_0 \approx 5\times10^{-4} P/\text{w}_0 $ is a small fraction, say $<$ 5\%=1/20 as a rule-of-thumb, of the total sample, the approximations used for the treatment of ionization and AC Stark Shift are justified. Turning this into a relation for a thermal sample, we obtain $u^2>20\,\overline{\text{v}_{\rm ion}}^2$, which corresponds to the relation $T > (1.7\times10^{-5} P/\text{w}_0)^2$. We can rewrite these relations, with the typical parameters used herein, as

\begin{subequations}
\begin{equation} \label{eq:vionvalid}
\text{v}_{\rm Lvalid} > 0.5~\text{m s}^{-1}\left(\frac{P}{200~\text{mW}}\right) \left(\frac{200~\mu\text{m}}{ \text{w}_0}\right) ,
 \end{equation}
\begin{equation} \label{eq:Tvalid}
   T_\text{validity} > 300~\mu\text{K} \left(\frac{P}{200~\text{mW}}\right)^2 \left(\frac{200~\mu\text{m}}{ \text{w}_0}\right)^2 .
 \end{equation}
  \end{subequations}
Based on studies of lines such as shown in Fig.~\ref{fig:gasinabox} we estimate the accuracy in the determination of the central frequency, without a magnetic trap, to be less than 20~Hz ($<  2$ parts in 10$^{14}$) under those conditions. In the case of a magnetic trap, as shown in Fig.~\ref{fig:CompareLineshapes},  with a field like $B_2(z)$ we obtain $<  7$ parts in 10$^{14}$ or even 0.2 parts in 10$^{14}$ in the case of a flatter field $B_1(z)$, in comparison with the OBE in the stated conditions. At these accuracies and higher, in the presence of high magnetic fields, a fully developed theory deriving relativistic corrections due to finite mass on the magnetic moment~\cite{PhysRevA.78.012504} justifies the necessary corrections to the magnetic moments when obtaining absolute frequency values.

The formalism presented could be extended to include the second-order Doppler effect ($\delta\nu/\nu =- (1/2) v^2/c^2$) which greatly affects experiments using cryogenic beams. For trapped samples the effect is small, of the order of parts in $10^{15}$ in the context of trapped atoms at 15~mK temperatures where $u/c \approx 5\times10^{-8}$.

\section{\label{sec:level7} Conclusions}
Using perturbation theory and further approximations we derive accurate analytical expressions for excitation lineshapes for hydrogen and antihydrogen laser spectroscopy on the 1S-2S transition. In the case of a magnetic trap the expressions need only a single integration along the laser axis taking into account the sample's density and frequency dependence on the trap inhomogeneous magnetic field. The formalism includes the AC Stark Shift and ionization by the laser, showing good accuracy as long as the number of atoms with low speeds, qualified in (\ref{eq:vionvalid}), is a small percentage of the total population. The use of these expressions allows for the study of many systematic effects -- such as magnetic field shape uncertainties, laser power, and sample deplenishing -- with trapped atoms.  The approximations employed result in a loss of accuracy by not capturing a small asymmetry (see Fig.~\ref{fig:DirectComparisonOBE}) with respect to the density matrix formalism and not taking fully into account the change of the atomic transition frequency as it crosses the laser beam in a gradient of the magnetic field in a trap (see Fig.~\ref{fig:CompareLineshapes}). These effects are estimated to affect the central frequency determination to parts in $10^{14}$ under the conditions considered when fitting well distributed points in the lineshape. Higher accuracies can be achieved by performing integrations (in $t_2, \text{v}_L, \rho_L$ and $z$) on (\ref{eq:c2numerical}), which is still quicker than computing grids with the OBE.  The present derivation, therefore, is suitable for analysis of the most accurate measurement with antimatter and its comparison with matter in a test of the CPT symmetry.

\section{\label{sec:level8} Acknowledgements}
The authors would like to acknowledge many useful discussions with the ALPHA 1S-2S spectroscopy task group analysing the 2023 data, and in particular Francis Robicheaux whose careful Monte-Carlo simulations provided invaluable insights and benchmarks into the excitation and detection process in the ALPHA experiment.

The authors acknowledge financial support from the Brazilian agencies CNPq, FAPERJ, and RENAFAE.

\appendix
\section{\label{sec:levelA} Factor $\alpha\approx\sqrt{\pi/6}$ }
In this appendix we treat the approximation that leads from (\ref{eq:c2Final}) to (\ref{eq:C2AC}).  By redefining variables, Eq.~\ref{eq:c2Final} can be written as

\begin{equation} \label{eq:c2FinalAppendix}
    c^{(2)}(t',\delta\omega) = \lambda\int_{-\infty}^{t'} \,dt\;e^{-i\left(\delta\omega t-\eta \left(1+{\rm Erf}(\xi t) \right)  \right)}e^{-\xi^2t^2} ,
\end{equation}

where
\begin{subequations}
  \begin{equation} \label{eq:AppendixSub1}
	\lambda=\frac{4\,P}{i\,\text{w}0^2}\left(2\sum{b}\kappa_{cba}\right)\;e^{-\frac{2\rho_{\rm L}^2}{\text{w}_0^2}},
  \end{equation}
  \begin{equation} \label{eq:AppendixSub2}
	\delta\omega=\delta\omega_{ca},
  \end{equation}
  \begin{equation} \label{eq:AppendixSub3}
	\eta=\frac{\sqrt{8\pi}P\:\beta_{\rm AC}}{\text{v}{\rm L}\:\text{w}_0}\exp\left[-\frac{2\rho{\rm L}^2}{\text{w}_0^2}\right],
  \end{equation}
and
  \begin{equation} \label{eq:AppendixSub4}
	\xi=\frac{\sqrt{2}\text{v}_\text{L}}{\text{w}_0}.
  \end{equation}
\end{subequations}

At resonance, $\delta\omega=0$, the integral in (\ref{eq:c2FinalAppendix}) has analytical solution for $t'\rightarrow\infty$,

\begin{equation} \label{eq:c2FinalAtResonanceAppendix}
    c^{(2)}(\infty,0) = \frac{\sqrt{\pi } e^{i \eta } \lambda  \sin (\eta )}{\eta  \xi } .
\end{equation}

Outside the resonance the integral is not analytical and we develop an approximation to obtain an analytical result. The error function can be expanded to first order in $t$, ${\rm Erf}(\xi t)\approx2\xi t/\sqrt{\pi}$. This approximation is good for $t\ll\sqrt{\pi}/(2\xi)$. The gaussian envelop in the integral has width $\sigma=1/(\sqrt{2}\xi)$. As  $\sigma$ is not much smaller than this value, $t\ll\sqrt{\pi}/(2\xi)$, the effect of the non-linearity of the error function, in the integral, is lost with this approximation. Since the most relevant condition for excitation happens at resonance, in order to achieve better accuracy in the integral result, one can multiply the expansion of the error function by a numerical factor $\alpha$ and equate the results given by the integral with the error function -- which, at $\delta\omega=0$, is analytical  -- and with this approximation. Simple algebra yields the result:

\begin{equation} \label{eq:NumericalFactorAppendix}
    \alpha(\eta)=\frac{1}{\eta}\sqrt{\pi\,{\rm Log}(\eta\,Csc(\eta))}.
\end{equation}

As seen above, there is a dependence of $\alpha$ with $\eta\propto \text{v}_\text{L}^{-1}$. For higher speeds, $\alpha$ can be expanded in power series of $\eta$:

\begin{equation} \label{eq:NumericalFactorSeriesAppendix}
    \alpha(\eta)=\sqrt{\frac{\pi}{6}}\left( 1+\frac{\eta^2}{60}\right) + \mathcal{O}(\eta^4).
\end{equation}

As long as $\eta^2/60\ll 1$, one can approximate $\alpha\approx\sqrt{\pi/6}$. For conditions typical of the ALPHA experiment, $\eta^2/60\approx0.01/\text{v}_\text{L}^2$. Even for atoms with $\text{v}_\text{L}=0.5 ~{\rm m~s}^{-1}$, $\eta^2/60<0.05$ and the approximation of $\alpha\approx\sqrt{\pi/6}$ is still good. As the detuning increases the approximation worsens but the $e^{-i\,\delta\omega\,t}$ term oscillates, quickly reducing the amplitude of the integral. A similar effect happens if the atom has high speeds and not only its spectrum will be broader (time of flight broadening) but the amplitude also decreases.

Out of resonance, with $\delta\omega\neq0$, the integral (\ref{eq:c2FinalAppendix}) performed using the approximation above is easily computed yielding:

\begin{equation} \label{eq:ApproxFinalAppendix}
     c^{(2)}_{\text{approx}}(\infty,\delta\omega)=\sqrt{\pi}\frac{\lambda}{\xi}e^{i\eta}\,e^{-\frac{\left(\delta\omega-\frac{2\alpha\xi\eta}{\sqrt{\pi}}\right)^2}{4\,\xi^2}}.
\end{equation}

The approximation is excellent at resonance, where the excitation rate is most relevant, but not so for low values of ($\text{v}_\text{L}\,\text{w}_0/P$) and away from the resonance peak. Due to this, the approximation fails to capture a subtle lineshape asymmetry which can cause a  small shift in the determination of the central peak. The level of accuracy, in comparison with the Optical Bloch Equations, is discussed in the text.  

\section{\label{sec:levelA} Magnetic Field Gradient }
In this appendix we discuss the magnetic field gradient as the atom crosses the laser beam. We can expand in first order the magnetic field near the crossing point, here called $z_c$, as $B(z) \approx B(z_c) +  (z-z_c) \beta$, where $\beta = \partial B(z)/\partial z$ at $z=z_c$  and which can be translated in a time dependence as $z = z_c + \text{v}_z t$. In this case, the frequency shift caused by the Zeeman and Diamagnetic terms can be written as

\begin{widetext}
  \begin{multline} \label{eq:Bgradient}
    \delta\omega_{B}(t) = \left[\,\delta\omega_{\text{Zee}}(B(z_c)) + \delta\omega_{\text{Dia}}(B(z_c))\,\right] + \frac{\beta\,\text{v}_z}{B(z_c)}\left[\delta\omega_{\text{Zee}}(B(z_c)) + 2\,\delta\omega_{\text{Dia}}(B(z_c))\right]\,t + \left( \frac{\beta\,\text{v}_z}{B(z_c)}\right)^2\delta\omega_{\text{Dia}}(B(z_c))\,t^2\\ 
    \equiv  \Delta\omega_B(B(z_c)) + \gamma_\beta{(z_c,\text{v}_z)}\,t + \Xi_\beta{(z_c,\text{v}_z)}\,t^2 ,
   \end{multline}
\end{widetext}
 with $\Delta\omega_B(B(z_c)) \equiv \delta\omega_{\text{Zee}}(B(z_c)) + \delta\omega_{\text{Dia}}(B(z_c))$, $\gamma_\beta{(z_c,\text{v}_z)}\equiv \beta\,\text{v}_z [\,\delta\omega_{\text{Zee}}(B(z_c)) + 2\,\delta\omega_{\text{Dia}}(B(z_c))\,]/B(z_c) $, and $\Xi_\beta{(z_c,\text{v}_z)} \equiv \left( \beta\,\text{v}_z/B(z_c)\right)^2\delta\omega_{\text{Dia}}(B(z_c))$.

With a reasonable magnetic field gradient ($\beta \lesssim 1\,\, \text{T/m}$) and $(\text{v}_z/\text{v}_\text{L})^2\ll1000$ and $\text{w}_0=200\,\,\mu m$, the term $\Xi_\beta{(z_c,\text{v}_z)}\,t^2$ can be neglected in the equation above. With this, the frequency shift seen by the atom can be linearized to the shift at the crossing point $\Delta\omega_B(B(z_c))$ plus a constant rate $\gamma_\beta{(z_c,\text{v}_z)}$ times $t$.

The time-varying frequency difference in $\delta\omega_{B}(t)$ can be added as $-\hbar\, \gamma_\beta{(z_c,\text{v}_z)} \,t\,\ket{a}\bra{a}$ in (\ref{eq:ACbraket}), resulting in an added term of form $-i\, \gamma_\beta{(z_c,\text{v}_z)} \,t^2/2$ in the exponential of (\ref{eq:a0}), the carries out to  the exponentials of (\ref{eq:c2b}), (\ref{eq:c2Final}), (\ref{eq:c2it}) and  (\ref{eq:c2numerical}). Notice that the analytical lineshape (\ref{eq:lineshapeIoni}) does not account for this time dependency of the atomic frequency, it just takes the frequency at the crossing point. 

Employing the approximations to the error functions in (\ref{eq:c2numerical}), it is possible to perform the analytical integral in $t_2$ with this linear rate of change of the atomic frequency, resulting in:

\begin{widetext}
  \begin{equation}\label{eq:C2ACIonBfield}
      \lvert C_{\rm B_{\text{field}}}(\delta\omega_{ca})\rvert^2=\frac{\gamma^2\,P^2}{\text{v}_{\rm L}^2 \text{w}_0^2\sqrt{1+\left(\frac{\gamma_\beta{(z_c,\text{v}_z)}}{2\text{v}_\text{L}^2\text{w}_0^2}\right)^2}} e^{-\frac{\text{w}_0^2}{4\text{v}_{\rm L}^2\left(1+\left(\frac{\gamma_\beta{(z_c,\text{v}_z)}}{2\text{v}_\text{L}^2\text{w}_0^2}\right)^2\right)}\left(\delta\omega{ca}-\Delta\omega_{\rm AC}\right)^2} e^{-2 \left(\frac{\sqrt{2 \pi} \beta_i\,P\,e^{-\frac{2\rho_{\rm L}^2}{\text{w}_0^2}}}{\text{v}_{\rm L} \text{w}_0}\right)}e^{-\frac{4\,\rho_{\rm L}^2}{\text{w}_0^2}},
\end{equation}
\end{widetext}
where the higher order ionization term $e^{\frac{1}{3}\left(\frac{\sqrt{2 \pi} \beta_i\,P\,e^{-\frac{2\rho_{\rm L}^2}{\text{w}_0^2}}}{\text{v}_{\rm L} \text{w}_0}\right)^2}$ that appears in (\ref{eq:C2ACion}) was already removed above.

This result can not be analytically integrated over $\text{v}_\text{L}f(\text{v}_\text{L})\,d\text{v}_\text{L}$ and over $g(\text{v}_z)d\text{v}_z$, where $g(\text{v}_z)$ is the thermal velocity distribution along the $z-\text{axis}$. However, one can see that the main effect of the magnetic field gradient as the atom crosses the laser is to cause a broadening and change in amplitude of excitation as a function of $\text{v}_\text{L}, \text{v}_\text{z}$, and $\beta$. Secondary effects, such as asymmetries, shifts, and oscillations, are lost due to the approximation done in the error functions. As discussed along the text, one can perform all integrals numerically, removing the ionization higher order term, to achieve better accuracy in a fitting procedure or for a more precise prediction of lineshape.

For the case of the OBE, to include the effect in first order of the magnetic field gradient one can simply substitute $\delta\omega_{ca}$ by $\delta\omega_{ca}(B(z_c))-\gamma_\beta{(z_c,\text{v}_z)}\,t$, with $\delta\omega_{ca}(B(z_c))=\delta\omega_{ca}-\Delta\omega_B(B(z_c))$. The change of magnetic field in the transverse direction is negligible because of the typical length of interaction being on the order of hundreds of $\mu$m (beam waist).

\bibliography{References.bib}

\end{document}